\documentclass[prd,twocolumn,nofootinbib]{revtex4}
\usepackage{graphicx, epsfig}
\usepackage{color}
\usepackage{mathrsfs}
\usepackage{bm}
\usepackage{amsmath}
\usepackage[stable]{footmisc}
\usepackage{ulem}

\newcommand{\be}{\begin{equation}}
\newcommand{\ee}{\end{equation}}
\newcommand{\ba}{\begin{eqnarray}}
\newcommand{\ea}{\end{eqnarray}}

\newcommand{\gapp}{\mathrel{\raise.3ex\hbox{$>$}\mkern-14mu
              \lower0.6ex\hbox{$\sim$}}}
\newcommand{\lapp}{\mathrel{\raise.3ex\hbox{$<$}\mkern-14mu
              \lower0.6ex\hbox{$\sim$}}}

\definecolor{bittersweet}{rgb}{1.0, 0.44, 0.37}
\definecolor{coolblack}{rgb}{0.0, 0.18, 0.39}
\definecolor{britishracinggreen}{rgb}{0.0, 0.26, 0.15}
\definecolor{coolgrey}{rgb}{0.55, 0.57, 0.67}
\definecolor{darkgreen}{rgb}{0.0, 0.2, 0.13}
\definecolor{darkmagenta}{rgb}{0.55, 0.0, 0.55}
\definecolor{eggplant}{rgb}{0.38, 0.25, 0.32}
\definecolor{fashionfuchsia}{rgb}{0.96, 0.0, 0.63}
\definecolor{ao(english)}{rgb}{0.0, 0.5, 0.0}

\begin{document}
\title{Dirac plus Nambu Monopoles in the Standard Model}
\author{George Lazarides$^{a}$, Qaisar Shafi$^{b}$, Tanmay Vachaspati$^{c}$}
\affiliation{
$^a$School of Electrical and Computer Engineering, Faculty of Engineering,
Aristotle University of Thessaloniki, Thessaloniki 54124, Greece \\
$^b$Bartol Research Institute, Department of Physics and Astronomy,
University of Delaware, Newark, DE 19716, USA\\
$^c$Physics Department, Arizona State University, Tempe, AZ 85287, USA.
}

\begin{abstract}
\noindent
We show how in the standard electroweak model three $SU(2)_L$ Nambu 
monopoles, each carrying electromagnetic (EM) and Z- magnetic fluxes, can 
merge (through Z-strings) with a single $U(1)_Y$ Dirac monopole to yield a composite monopole 
that only carries EM magnetic flux. Compatibility with the Dirac quantization condition requires this 
composite monopole to carry six quanta ($12 \pi /e$) of magnetic charge, independent of the electroweak 
mixing angle $\theta_w$. The Dirac monopole 
is not regular at the origin and the energy of the composite monopole is therefore divergent. We discuss 
how this problem is cured by embedding $U(1)_Y$ in a grand unified group such as $SU(5)$. A second
composite configuration with only one Nambu monopole and a colored $U(1)_Y$ Dirac monopole that has
minimal EM charge of $4\pi/e$ is also described.
Finally, there exists a configuration with an EM charge of $8\pi/e$ as well as screened
color magnetic charge.
\end{abstract}

\maketitle

\section{Introduction}
\label{introduction}

It is widely recognized that the Standard Model (SM) based on the gauge group 
$SU(3)_c \times SU(2)_L \times U(1)_Y/Z_3\times Z_2$ does not contain topologically 
stable finite energy monopoles. 
This stems from the fact that the vacuum manifold
in the SM is a three-sphere and does not have a non-trivial
second homotopy group that is required for the existence of  topological 
finite  energy monopoles.
Note that this result is 
unaffected if additional scalar and/or matter fields are introduced in the SM. 
Nevertheless, as we show, interesting composite
topological structures are present in the SM and its extensions such as grand
unified theories (GUTs).

It was recognized sometime ago~\cite{Nambu:1977ag,Vachaspati:1992fi,Achucarro:1999it} 
that a 't Hooft-Polyakov type~\cite{HOOFT1974276,Polyakov:1974ek} monopole configuration 
can be realized within the $SU(2)_L$ sector of the SM. This is achieved by utilizing a Higgs doublet 
field configuration which effectively behaves like a scalar $SU(2)_L$ triplet with 
hypercharge $Y=0$. This yields a 
monopole configuration that carries one unit of $T_L^ 3$ magnetic charge, where 
$T_L^3={\rm diag}(1,-1)$ denotes the 
third generator of $SU(2)_L$. 
In other words, the monopole emits both electromagnetic (EM) 
and Z magnetic flux, where the latter appears in a flux tube (or Z string). The 
EM and Z magnetic flux 
carried by the monopole are found to be $(4\pi/e) \sin^{2}\theta_w$ 
and $(4\pi/e) \sin\theta_w\cos\theta_w$  respectively
where $\theta_w$ is the electroweak mixing angle with $\sin^2\theta_w =0.23$.
This monopole cannot be purely electromagnetic in nature with 
the Z flux screened within its core (size $\sim M_Z^{-1}$) 
because it would violate Gauss's
law pertaining to the magnetic component of $U(1)_Y$. 

A recent paper~\cite{PhysRevD.103.095021} shows that the Nambu monopole configurations 
routinely arise in GUTs, with the simplest example provided by SU(5). A new feature here is the 
presence of an elementary scalar $SU(2)_L$ triplet with $Y=0$ that acquires an induced vacuum 
expectation value (VEV) from the electroweak breaking. In addition, the quantized value of the 
Z flux is also predicted by exploiting the well-known prediction of $\sin^2 \theta_w=3/8$.

Motivated by these considerations, it seems appropriate to enquire whether a purely 
electromagnetic monopole configuration can be realized in the SM.
Clearly, this cannot happen without involving the $U(1)_Y$ sector, 
in contrast to the claim in Refs.~\cite{Hung:2020vuo,Ellis:2020bpy}.
Ignoring 
the singular nature of a Dirac monopole~\cite{dirac}
for a moment,  it turns out that a purely  electromagnetic monopole can be realized by 
combining via Z strings a single Y monopole of appropriate magnetic charge with three 
Nambu monopoles. This composite monopole structure carries a magnetic charge of ($12 \pi / e$) in 
order to satisfy the Dirac quantization conditions in the presence of quarks and charged leptons. All 
dependence on $\sin^2 \theta_w$ disappears in arriving at this conclusion.
We show how the embedding of $U(1)_Y$ in a grand
unified framework such as SU(5) resolves the singular
nature of the above composite monopole.
As expected, the monopole 
turns out to be superheavy with mass comparable to the GUT scale.
Two additional examples are provided of configurations that carry EM charges
of $4\pi/e$ and $8\pi/e$ as well as screened color magnetic charge.

\section{Combining Nambu and Dirac Monopoles in the Standard Model}
\label{main body}

We are admitting the possibility of a $U(1)_Y$ Dirac monopole in the SM. From the
Dirac quantization condition such a monopole carries magnetic hypercharge
\be
m_Y 
= \frac{12\pi}{g'} n, \ \ n\in {\cal Z},
\ee
where $g'$ is the hypercharge gauge coupling constant,
and the left-handed quark doublet $(u,d)_L$ has hypercharge $1/6$.
Next consider a Nambu monopole in the symmetry breaking $SU(2)_L \to U(1)_L$.
Such a monopole has weak magnetic charge given by the condition
\be
m_L = \frac{2\pi}{g/2} n' = \frac{4\pi}{g} n', \ \ n' \in {\cal Z},
\ee
where $g$ is the $SU(2)_L$ coupling constant. We can work in the unitary gauge for the
$SU(2)_L$ monopole such that the magnetic flux from the monopole consists only of 
$W^3_\mu$ gauge field, where $W^a_\mu$ denote the usual $SU(2)_L$ gauge fields.

We now wish to put the Nambu $SU(2)_L$ and Dirac $U(1)_Y$ monopoles together and re-express
the charges in terms of the post-electroweak symmetry breaking gauge fields: $Z_\mu$
and $A_\mu$. 
The relation between $(W^3_\mu,Y_\mu)$ 
and $(Z_\mu,A_\mu)$ is given by 
\ba
W^3_\mu = \cos\theta_w Z_\mu + \sin \theta_w A_\mu , \\
Y_\mu = \cos\theta_w A_\mu - \sin\theta_w Z_\mu.
\ea
Therefore, the $Z_\mu$ and $A_\mu$ magnetic charges on $n'$ $SU(2)_L$ monopoles are
\be
m_{LZ} = \frac{4\pi n'}{g} \cos\theta_w, \ \
m_{LA} = \frac{4\pi n'}{g} \sin\theta_w,
\ee
and on $n$ $U(1)_Y$ monopoles are
\be
m_{YZ} = - \frac{12\pi n}{g'} \sin\theta_w, \ \
m_{YA} = \frac{12\pi n}{g'} \cos\theta_w.
\ee

The net $Z_\mu$ and $A_\mu$ magnetic charges on a conglomerate of $n$ $U(1)_Y$ and 
$n'$ $SU(2)_L$ monopoles are
\ba
m_{Y+L,Z} &=&  \frac{4\pi n'}{g} \cos\theta_w- \frac{12\pi n}{g'} \sin\theta_w, \\
m_{Y+L,A} &=& \frac{4\pi n'}{g} \sin\theta_w+ \frac{12\pi n}{g'} \cos\theta_w.
\label{mYLA}
\ea

This configuration should not
have any net $Z$ magnetic charge because the $Z$ magnetic fields are confined
once the electroweak symmetry is broken. Any net $Z$ flux would form a string 
that would confine the monopole conglomerate to an anti-conglomerate. Thus, we require 
\be
 \frac{4\pi n'}{g} \cos\theta_w- \frac{12\pi n}{g'} \sin\theta_w =0.
 \label{Zcondition}
 \ee
 Together with the relations
 \be
 \cos\theta_w = \frac{g}{\sqrt{g^2+ g'^2}}, \ \ 
 \sin\theta_w = \frac{g'}{\sqrt{g^2+ g'^2}},
 \label{thetawg}
 \ee
 the constraint in \eqref{Zcondition} gives
 \be
 n' = 3 n,
 \label{n3n}
 \ee
 and so the conglomerate should contain three times as many Nambu monopoles
 as the Dirac Y-monopoles.
 
 Next let us consider the EM magnetic charge on the conglomerate.
 From \eqref{mYLA}, taking into account \eqref{thetawg} and \eqref{n3n}, we get
 \be
m_{Y+L,A} = \frac{12\pi}{e} n 
 \ee
 where $e= gg'/\sqrt{g^2+g'^2}$ is the electromagnetic coupling constant. Therefore
 the minimal EM charge of the conglomerate is $6 \times 2\pi /e$ and
 this result is independent of $\sin^2\theta_w$. 

\begin{figure}
      \includegraphics[width=0.48\textwidth,angle=0]{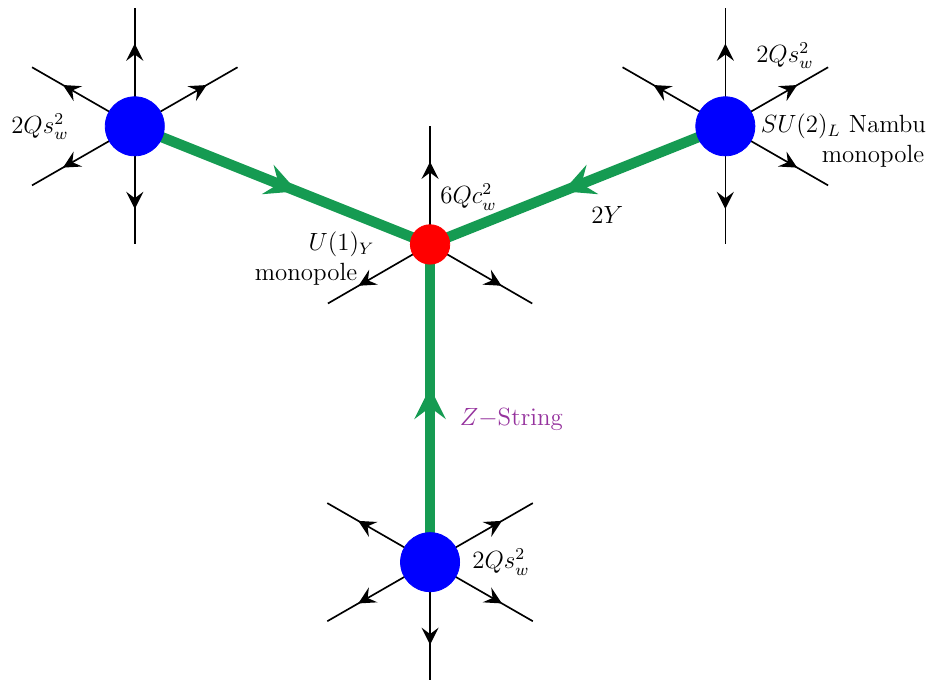}
        \caption{
      A $U(1)_Y$ Dirac monopole (red) with minimal EM magnetic flux $6Q c_w^2$,
where $c_w=\cos\theta_w$, gets attached to three Nambu $SU(2)_L$ monopoles
(blue) each with EM magnetic flux $2Qs_w^2$, where $s_w=\sin\theta_w$.
The Z-component of the $SU(2)_L$ flux of the Nambu monopoles 
is confined in strings with flux $T^3_L-2s^2_w Q$
that connect them to the $U(1)_Y$ Dirac monopole. 
The conglomerate yields
an electromagnetic monopole that carries six quanta ($12\pi/e$) of EM
magnetic charge.
}
  \label{monopole}
\end{figure}

In Fig.~\ref{monopole} we sketch the minimal allowed monopole conglomerate.
From (5), we see that the minimal ($n'=1$) magnetic flux
along the Z-string is $4\pi/g_z$, where $g_z=g/c_w$ is the
gauge coupling associated with the Z gauge boson with $c_w=\cos\theta_w$. This
string corresponds to a $4\pi$ rotation about the corresponding
generator $T^3_L/2-s^2_wQ$ and therefore it carries magnetic flux
$T^3_L-2s^2_wQ$, where $s_w=\sin\theta_w$. The minimal Nambu monopole also carries
EM magnetic charge $(4\pi/g)s_w=(4\pi/e)s^2_w$
corresponding to magnetic flux $2Qs^2_w$.

In Sec.~\ref{colored} we show how 
a monopole configuration with EM charge $4\pi/e$ is realized by merging a single
Nambu monopole with a {\it colored} $U(1)_Y$ monopole.

\section{Nambu and $U(1)_Y$  Monopoles in $SU(5)$}
\label{discussion}
The embedding of the SM in a grand unified framework has important ramifications, 
especially as far as our current discussion is concerned. For instance, in minimal SU(5) 
the Nambu monopole contains an inner core due to the presence of a heavy $SU(2)_L$ 
scalar triplet field that acquires an induced VEV through 
mixing with the SM Higgs doublet. 
The Nambu monopole carries
$(3/4) ( 2 \pi /e)$ unit of Dirac magnetic charge, and the Z magnetic flux is also predicted [6].
The composite electromagnetic monopole discussed above is now realized by considering 
a $U(1)_Y$ monopole with winding number six, which appears from the symmetry breaking
$SU(5) \xrightarrow[]{} SU(3)_c \times SU(2)_L \times U(1)_Y/Z_3\times Z_2$.  At this stage 
the monopole has 
a core size of order $M_{GUT}^{-1}$ and mass around $10\;M_{GUT}$. Clearly, the singularity 
encountered with such a monopole in the SM setting has now disappeared.

Following electroweak breaking the monopole carries an appropriate combination of EM
and Z magnetic fluxes, with the latter in the form of Z-strings. In order to retain a finite energy 
configuration this $U(1)_Y$ monopole in $SU(5)$ must merge with three Nambu monopoles 
to yield a composite structure that carries six quanta of Dirac magnetic charge (Fig.~1).
Note that in the absence of quarks just two quanta $(4 \pi / e)$ would have been compatible 
with the Dirac quantization condition. 
Indeed, in the $SU(5)$ breaking to $SU(3)_c \times SU(2)_L \times U(1)_Y/Z_3\times Z_2$ the fundamental 
topologically stable monopole carries $SU(3)_c$, $SU(2)_L$ and $U(1)_Y$ magnetic charges. 
The subsequent breaking of the SM symmetry to $SU(3)_c \times U(1)_{em}/Z_{3}$ yields a 
monopole carrying a single quantum $(2\pi/e)$ of Dirac magnetic charge as well as screened 
color magnetic charge. Quark confinement in this case is
required in order for the Dirac quantization condition to be
satisfied beyond the screening radius $\sim \Lambda_{QCD}^{-1}$.

Finally, let us note that a purely electromagnetic monopole carrying three quanta of Dirac 
magnetic charge and accompanied by no other field exists in SU(5), and a classical solution 
has been shown in Ref.~\cite{RUBAKOV1983240}.

\section{Colored Dirac Monopoles}
\label{colored}

\begin{figure}
      \includegraphics[width=0.48\textwidth,angle=0]{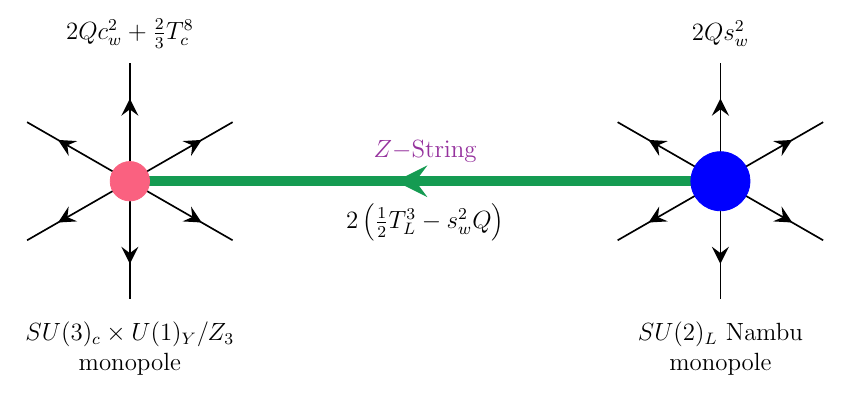}
        \caption{
        A doubly charged colored Dirac monopole (pink)
        and a Nambu monopole (blue) connected by a
Z-string with magnetic flux $T^3_L-2s^2_wQ$. The total flux
of this configuration is $2Q+2T^8_c/3$ of which the colored part
will be screened due to QCD effects.
}
  \label{su3u1case}
\end{figure}

The smallest homotopically non-trivial loop in the SM gauge group
$SU(3)_c\times SU(2)_L\times U(1)_Y/Z_3\times Z_2$  consists of a $2\pi$ rotation 
about $Y$ -- for the left-handed quark doublet $Y=1/6$  -- accompanied by a $\pi$
rotation about $T^3_L={\rm diag}(1,-1)$ and a $2\pi/3$ rotation about $T^8_c={\rm diag}(1,1,-2)$.
The minimal singular Dirac monopole then has magnetic flux $Y+T^3_L/2+
T^8_c/3$. The magnetic flux of the doubly charged Dirac monopole can
be written as $2Y+2T^8_c/3$ since the loop in $SU(2)_L$ becomes
homotopically trivial and can be removed. After electroweak symmetry
breaking, the colored Dirac monopole can absorb the Z-flux 
emerging from the Nambu monopole (Fig.~\ref{su3u1case})
(see also Ref.~\cite{Preskill:1992bf}).
The total EM plus color flux of the conglomerate is
$2Q+2T^8_c/3$ with the color part screened for distances larger than
the QCD scale.

\section{Structure and stability}
\label{stability}

Figs.~\ref{monopole} and \ref{su3u1case} show the purely EM monopoles in terms of
their constituent Dirac and Nambu monopoles. However, we do not expect these 
diagrams to realistically show classical solutions of the field theory. For example,
the colored monopole of Fig.~\ref{su3u1case} is more likely to be a structureless, 
point-like $SU(3)\times U(1)_Y/Z_3$ Dirac monopole surrounded by a cloud 
of electroweak fields that make up the Nambu monopole. 
Other solutions have been proposed in 
Ref.~\cite{Cho:1996qd}
though these solutions do not correspond to the correctly quantized monopoles
depicted in Figs.~\ref{monopole} and \ref{su3u1case} -- the conglomerate in 
Fig.~\ref{monopole} has three (not one) Nambu monopoles, and the conglomerate
in Fig.~\ref{su3u1case} has net color charge. 

If the Dirac monopoles are regularized
as $SU(5)$ monopoles, Fig.~\ref{su3u1case} corresponds to the winding 2 monopole
resulting from the symmetry breaking to $SU(3)_c\times SU(2)_L\times U(1)_Y/Z_3\times Z_2$,
while Fig.~\ref{monopole} corresponds to the winding 6 grand unified monopole.
The odd winding $SU(5)$ monopoles carry $SU(2)_L$ magnetic charge and once
electroweak symmetry breaks, two odd winding monopoles combine to result
in an $SU(2)_L$ neutral monopole~\cite{Ng:2008mp}. Further, it has been shown that 
the $SU(5)$ monopole
with winding $n=5$ is unstable to fragmentation into $n=2+3$ monopoles, and 
monopoles with winding $n \ge 7$ are unstable to fragmentation into $n=6+(n-6)$
monopoles~\cite{Vachaspati:1995yp,Liu:1996ea}.
Then the only other relevant winding to consider in the $SU(5)$ regularization 
is $n=4$. This monopole will connect with two Nambu monopoles to form a 
conglomerate that carries color and EM magnetic charges but the color magnetic
field will get screened to give a pure EM monopole with flux $4Q$ 
(EM charge $8\pi/e$)
 at large distances.

\section{Summary}
\label{summary}

We have explored how purely EM monopole configurations may arise in the 
SM. We do not expect such configurations to have well defined energy 
because of the singular nature of the Dirac monopole associated with the $U(1)_Y$ sector of
 the electroweak model. Barring this, we identified a configuration consisting of a single $U(1)_Y$ 
 Dirac monopole which is linked via Z strings to three Nambu monopoles. This composite monopole 
 carries six units ($12 \pi / e$) of EM magnetic charge in order to be compatible with the Dirac 
 quantization condition in the presence of quarks. Although the Y and Nambu monopoles both carry 
 $\theta_w$ dependent EM and Z magnetic charges, the end result leaves $\theta_w$ 
 undetermined in the $SU(2)_L \times U(1)_Y/Z_2$ framework. We also show how this singular composite 
 monopole gets regularized and obtains a finite, albeit superheavy mass, from embedding the SM in a 
 GUT. In the second example a colored $U(1)_Y$ monopole merges
 with a single Nambu monopole to yield a composite topological structure carrying an EM
 magnetic charge of $4\pi/e$ as well as screened color magnetic charge.
A third example consists of two Nambu monopoles and a single colored Dirac $U(1)_Y$ monopole
and has EM magnetic charge $8\pi/e$.

Let us recall here that models such as 
$SU(4)_c \times SU(2)_L  \times SU(2)_R$~\cite{Pati:1974yy} and 
$SU(3)_c \times SU(3)_L \times SU(3)_R$ predict electric charge 
quantization~\cite{Lazarides:2019xai,Lazarides:2021tua}, but gauge coupling unification 
in these models is not automatic. 
In this case the masses of the composite monopoles can lie in the multi-TeV range.
It is worth stressing that these two models also predict the existence of topologically stable magnetic 
monopoles that respectively carry two and three Dirac quanta of magnetic charge, and which may be 
accessible at high energy colliders.

Before concluding we offer some remarks regarding the observability of these composite monopoles at the 
LHC and future high energy colliders. The production cross section of a composite coherent quantum state 
including these composite monopole states is expected to be exponentially suppressed in Drell-Yan 
processes involving particle collisions~\cite{Mavromatos:2020gwk}. This is reminiscent of the exponential 
suppression in quantum 
mechanical tunneling. It has been argued that the magnetic analogue of the Schwinger mechanism could 
be exploited to significantly enhance the production cross sections of these extended structures and propel 
them in the observable range. With adequately strong magnetic fields the (dual) Schwinger mechanism may 
lead to an observable cross section for monopole pair production in heavy ion collisions~\cite{Gould:2021bre}. 
The MoEDAL experiment has recently presented the first results~\cite{Acharya:2021ckc} based 
on this idea.

For the composite monopoles to be accessible at high energy colliders their mass should lie in the 
TeV range or so.  This means that the $U(1)_Y$ component of the electroweak sector of the SM should be
embedded in a unifying gauge symmetry such as $SU(3)_L \times SU(3)_R$, which may be broken at the 
TeV scale if we do not require gauge coupling unification of the left-right $SU(3)$ 
symmetries~\cite{Lazarides:2021tua,Kephart:2017esj}. For a discussion of monopoles in cosmic 
rays, see Ref.~\cite{Kephart:2017esj} and other references listed there.

\vspace{0.5cm}
\noindent 
{\bf Acknowledgments.}\,
{We thank Amit Tiwari for drawing the figures and P.Q. Hung and Nick Mavromatos for 
discussions.
This work is supported by the Hellenic Foundation for Research 
and Innovation (H.F.R.I.) under the ``First Call for H.F.R.I. 
Research Projects to support Faculty Members and Researchers and 
the procurement of high-cost research equipment grant'' (Project 
Number: 2251). Q.S. is supported in part by the DOE Grant 
DE-SC-001380.
T.V. is supported by the U.S. Department of Energy, Office of High Energy 
Physics, under Award DE-SC0019470 at ASU.}

%\bibliographystyle{unsrt}
%\bibstyle{aps}
%\bibliography{paper.bib}

\end{document}